\documentclass[conference]{IEEEtran}
\usepackage{amsmath,amssymb,amsfonts}
\usepackage{graphicx}
\usepackage{textcomp}
\usepackage{xcolor}
\usepackage{booktabs}
\usepackage{threeparttable}
\usepackage{adjustbox}
\usepackage{algorithm}
\usepackage{algorithmic}

\IEEEoverridecommandlockouts

\begin{document}
\setlength{\IEEEilabelindent}{0pt}

\title{Parameter-Efficient CT Reconstruction via Deep Graph Laplacian Regularization}

\author{
\IEEEauthorblockN{Veera Varuni Radhakrishnan\textsuperscript{1},
Chinthaka Dinesh\textsuperscript{1},
Qurat-ul-Ain Azim\textsuperscript{1}}
\IEEEauthorblockA{
\textsuperscript{1}\textit{Mechanical and Industrial Engineering Department}\\
\textit{Northeastern University}\\
Vancouver, Canada\\
\{radhakrishnan.ve, c.herathgedara, q.azim\}@northeastern.edu}
}
\maketitle

\begin{abstract}
Low-dose computed tomography (LDCT) reconstruction faces a critical 
tradeoff between reconstruction quality and resource requirements. 
While recent deep learning methods achieve state-of-the-art performance, they 
typically rely on over 500,000 parameters trained on large-scale datasets exceeding 35,000 scans. This work investigates whether graph-based 
regularization can provide meaningful noise reduction under strict 
resource constraints. We propose Deep Graph Laplacian Regularization 
(Deep GLR), integrating quadratic graph regularization into a Proximal 
Forward-Backward Splitting optimization framework with three 
lightweight CNN modules. Evaluated on the LoDoPaB-CT benchmark, Deep GLR achieves 
30.70 dB PSNR, representing a 6.33 dB improvement over filtered backprojection, while using 
only 91,848 parameters trained on 1000 samples (2.8\% of standard 
training set). Compared to benchmark methods, this represents 5.8 times 
better parameter efficiency and 30 times better data efficiency per dB 
improvement. The learned graph bandwidth parameter ($\epsilon$=1.25) converges 
to interpretable values, suggesting the method captures meaningful 
image priors rather than overfitting. While a 13 dB gap remains versus 
state-of-the-art methods, results demonstrate that graph-based 
regularization provides a favorable efficiency-quality tradeoff for 
resource-constrained medical imaging scenarios.
\end{abstract}


\section{Introduction}

Low-dose computed tomography (LDCT) reconstruction addresses the major challenge of reducing patient radiation exposure while maintaining diagnostic image quality. Studies show measurable cancer risks from CT radiation exposure, with excess relative risk estimates of 1.05 per 100 mGy for leukemia and 0.80 per 100 mGy for brain tumors~\cite{little2021epidemiological}. With approximately 6.4 million CT scans performed annually in Canada alone~\cite{cadth2024canadian}, radiation dose reduction is essential for patient safety going forward.

The CT reconstruction problem involves recovering the attenuation coefficient distribution $\mathbf{x} \in \mathbb{R}^n$ from noisy projection measurements $\mathbf{y} \in \mathbb{R}^m$:
\begin{equation}
\mathbf{y} = \mathbf{A}\mathbf{x} + \mathbf{n}
\label{eq:forward_model}
\end{equation}
where $\mathbf{A}$ represents the discretized Radon transform and $\mathbf{n}$ denotes Poisson-distributed measurement noise.

While recent deep learning methods achieve impressive reconstruction 
quality \cite{shen2024deep, ding2019low}, they face two practical barriers in medical imaging 
tasks \cite{esteva2019guide}: (1) requirement for large labeled training sets, which may be unavailable due to privacy 
constraints, rare pathologies, or limited institutional resources, and 
(2) models with a large number of parameters requiring substantial computational 
resources for both training and clinical deployment. These constraints 
motivate investigation of whether classical optimization frameworks 
enhanced with lightweight learning can achieve acceptable reconstruction 
quality under strict resource limitations.

Traditional analytical reconstruction methods like filtered backprojection (FBP) provide computational efficiency but also cause severe noise production in low-dose scenarios. Recent deep learning approaches have achieved significant improvements, with convolutional neural network (CNN) methods that use millions of parameters and extensive training. However, these approaches often lack interpretability that can benefit medical applications.

This work investigates whether graph-based regularization integrated with 
classical optimization can achieve meaningful noise reduction under strict 
resource constraints. Our specific contributions are as follows:
\begin{enumerate}
\item Adaptation of Deep Graph Laplacian  Regularization (Zeng et al. \cite{zeng2019deep}) to CT 
   reconstruction by integrating quadratic graph regularization into the 
   Proximal Forward-Backward Splitting framework (Ding et al. \cite{ding2019low}), enabling 
   parameter-efficient reconstruction with a smaller number of parameters and closed-
   form gradients.

\item Demonstration that this approach achieves clear performance improvement over 
   filtered backprojection while using approximately one-fifth the 
   parameters of existing deep learning methods, with effective training 
   on limited samples.

\item Analysis of learned parameter behavior showing convergence to 
   interpretable values, indicating that the framework captures meaningful image priors rather than overfitting to the training data.

\item Practical implementation and evaluation on the LoDoPaB-CT 
   benchmark \cite{leuschner2021lodopab}, and providing a reproducible 
   baseline for investigating efficiency-quality tradeoffs in resource-constrained CT reconstruction.
\end{enumerate}
\section{Related Work}

\subsection{CT Reconstruction Methods}
Filtered backprojection is a traditional method used for clinical CT reconstruction, which involves analytical inversion of the Radon transform through frequency domain filtering. As demonstrated in the LoDoPaB-CT benchmark, FBP achieves rapid reconstruction but yields $30.52\pm3.10$ dB PSNR on low-dose test data due to direct noise propagation from the projection  ~\cite{leuschner2021lodopab}. The method's linear nature preserves the quantum noise present in low-photon measurements, resulting in characteristic streak artifacts that compromise the diagnostic quality.

Iterative reconstruction approaches address these limitations by treating image recovery as constrained optimization problems. These methods incorporate explicit regularization terms such as total variation, sparsity constraints, or smoothness priors to tackle the ill-posed inverse problem. However, the effectiveness of such approaches depends heavily on the appropriateness of these handcrafted priors for medical image characteristics, often requiring manual parameter tuning for different imaging protocols, which is not ideal.

\subsection{Deep Learning for CT Reconstruction}
Recent deep learning approaches can be categorized into post-processing methods and algorithm unrolling approaches. Post-processing methods apply CNN denoising to FBP reconstructions but treat the reconstruction process as separate from the physics of CT acquisition. FBPConvNet and similar approaches achieve improved image quality but lack any integration with the underlying inverse problem formulation.


Algorithm unrolling methods \cite{monga2021algorithm} address these limitations by combining deep 
learning with iterative optimization. Ding et al. \cite{ding2019low} developed Proximal 
Forward-Backward Splitting (PFBS) methods that unroll proximal forward-
backward splitting with data-driven regularization, demonstrating superior 
performance over both conventional analytical methods and pure CNN post-
processing approaches. However, their approach replaces explicit 
regularization terms with learned operators, increasing parameter count 
to 500K+ to capture complex image priors

Our work diverges by maintaining explicit mathematical regularization 
(Graph Laplacian) rather than replacing it with learned 
operators. This design choice trades maximum flexibility for parameter 
efficiency and interpretability.

\subsection{Graph-Based Image Regularization}
Graph signal processing treats image pixels as nodes in similarity graphs, enabling sophisticated regularization schemes. Vu et al.~\cite{vu2021unrolling} demonstrated effective image denoising using Graph Total Variation (GTV), which employs $L_1$-norm regularization requiring iterative thresholding operations. In contrast, Zeng et al.~\cite{zeng2019deep} introduced Deep Graph Laplacian Regularization (Deep GLR) using quadratic formulation that provides closed-form gradients and computational stability.

The choice between $L_1$ and $L_2$ graph regularization involves fundamental 
tradeoffs. Vu et al. ~\cite{vu2021unrolling} demonstrated that Graph Total Variation ($L_1$-
norm) effectively preserves edges in denoising tasks. However, the non-
smooth formulation requires proximal operators (soft thresholding), 
complicating integration with automatic differentiation frameworks. On the other hand, Zeng et al. ~\cite{zeng2019deep} employed Deep Graph Laplacian (L2-norm) for natural 
image denoising, showing that quadratic formulations enable stable end-
to-end training despite over-smoothing tradeoffs. Their work focused on 
additive Gaussian noise in natural images.

We adapt Zeng's framework to CT reconstruction, where challenges differ 
substantially: (1) Poisson noise in projection domain rather than 
Gaussian noise in image domain, (2) ill-posed inverse problem requiring 
forward model integration rather than direct denoising, and (3) medical 
image characteristics (piecewise constant tissue regions) versus natural 
image statistics. Our integration of Deep GLR with PFBS addresses these 
domain-specific requirements.


\section{Methodology}

\subsection{Mathematical Framework}

The LDCT reconstruction problem \eqref{eq:forward_model} is ill-posed due to noise 
amplification. We formulate 
reconstruction as a regularized inverse problem that balances data 
fidelity with image smoothness constraints. The regularization term 
should: (1) provide noise reduction, (2) preserve anatomical boundaries, 
and (3) enable efficient gradient computation for deep learning 
integration.

LDCT reconstruction can be formulated as the regularized optimization problem:
\begin{equation}
\min_{\mathbf{x}} \frac{1}{2}\|\mathbf{A}\mathbf{x} - \mathbf{y}\|_2^2 + \mu(\mathbf{x}^T \mathbf{L} \mathbf{x})
\label{eq:optimization}
\end{equation}
where $\mathbf{L} \in \mathbb{R}^{n \times n}$ denotes the graph Laplacian matrix, $\mu > 0$ is the learned regularization parameter and $\mathbf{A} \in \mathbb{R}^{m \times n}$ is the discrete Radon transform operator mapping the image domain ($n = 362 \times 362$ pixels) to the sinogram domain ($m = 1000 \times 513$ detector measurements), implemented via the ODL ray transform~\cite{leuschner2021lodopab}. The Graph Laplacian is defined as $\mathbf{L} = \mathbf{D} - \mathbf{W}$, where $\mathbf{W}$ is the weighted adjacency matrix and $\mathbf{D}$ is the diagonal degree matrix.

The quadratic Graph Laplacian Regularization term:
\begin{equation}
\text{GLR}(\mathbf{x}) = \mathbf{x}^T \mathbf{L} \mathbf{x} = \sum_{(i,j) \in E} w_{ij}(x_i - x_j)^2,
\label{eq:glr}
\end{equation}
where $E$ is the edge set of the graph, provides closed-form gradient $\nabla \text{GLR}(\mathbf{x}) = 2\mathbf{L}\mathbf{x}$, enabling stable optimization without iterative operators required by $L_1$-norm methods.

\subsection{Network Architecture}
The Deep GLR architecture consists of three CNN modules integrated within the PFBS framework:

\textbf{1) Pre-filtering Network (CNN$_{Yb}$)}: A 4-layer convolutional network with residual connections performing initial noise reduction. Architecture: [1$\rightarrow$32$\rightarrow$32$\rightarrow$32$\rightarrow$1] channels with 3$\times$3 kernels and BatchNorm+ReLU activations.

\textbf{2) Feature Extraction Network (CNN$_F$)}: A 5-layer network learning $K=3$ dimensional features per pixel. Architecture: [1$\rightarrow$16$\rightarrow$32$\rightarrow$32$\rightarrow$16$\rightarrow$3] channels with 5$\times$5 kernels. Output: $\mathbf{f} \in \mathbb{R}^{3 \times H \times W}$ feature tensor.

\textbf{3) Parameter Learning Network (CNN$_\mu$)}: This layer outputs adaptive regularization weight $\mu \in [0, 0.1]$ per image via 2 conv layers + global pooling + fully connected layer with Sigmoid$\times$0.1 constraint.

\textbf{4) Post-processing Network}: Final refinement network with residual connection for artifact removal.

\textbf{Learnable Parameters}: The architecture includes CNN weights $\theta = \{\theta_{Yb}, \theta_F, \theta_\mu\}$ and two scalar parameters: step size $\alpha \in [0, 0.1]$ and graph bandwidth $\varepsilon \in [1.0, 1.5]$, with constraints enforced via sigmoid activations and parameter clamping during training.

\subsection{Graph Construction}
Edge weights between pixels $i$ and $j$ are computed using learned features within an 8-connected neighborhood topology. For each pixel $i$, edges are established only with its 8 adjacent neighbors (horizontal, vertical, and diagonal), creating a sparse graph structure with fixed connectivity pattern.

Given feature tensor $\mathbf{f} \in \mathbb{R}^{3 \times H \times W}$ from CNN$_F$, edge weights are:
\begin{equation}
w_{ij} = \exp\left(-\frac{\|\mathbf{f}_i - \mathbf{f}_j\|_2^2}{2\varepsilon^2}\right)
\label{eq:edge_weights}
\end{equation}
where $\mathbf{f}_i \in \mathbb{R}^3$ represents the feature vector at pixel location $i$. The resulting adjacency matrix $\mathbf{W} \in \mathbb{R}^{n \times n}$ where $n = H \times W = 131,044$ contains non-zero entries only for neighboring pixel pairs, yielding the sparse graph Laplacian $\mathbf{L} = \mathbf{D} - \mathbf{W}$.

\subsection{PFBS Algorithm Integration}
The PFBS framework alternates between data fidelity and regularization updates. Each iteration $k$ performs:
\begin{align}
\mathbf{x}^{k+1/2} &= \mathbf{x}^k - \alpha\nabla_{\text{data}}(\mathbf{x}^k) \label{eq:pfbs_forward}\\
\mathbf{x}^{k+1} &= \mathbf{x}^{k+1/2} - \alpha\mu\nabla_{\text{GLR}}(\mathbf{x}^{k+1/2}) \label{eq:pfbs_backward}
\end{align}
where $\nabla_{\text{data}}(\mathbf{x}) = 2\mathbf{A}^T(\mathbf{A}\mathbf{x} - \mathbf{y})$ and $\nabla_{\text{GLR}}(\mathbf{x}) = 2\mathbf{L}\mathbf{x}$.

\begin{figure*}[!t]
\centering
\includegraphics[width=\linewidth]{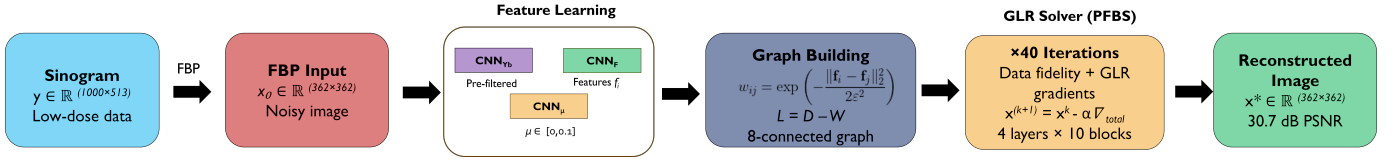}
\caption{Deep GLR pipeline showing integration of CNN feature learning with PFBS optimization for CT reconstruction. The method combines pre-filtering, graph construction, and iterative optimization in an end-to-end trainable framework with 40 total iterations across 4 layers}
\label{fig:pipeline}
\end{figure*}

The complete architecture uses 4 GLR layers with 10 blocks each, totaling 40 PFBS iterations with shared CNN weights across all blocks. Fig.~\ref{fig:pipeline} illustrates the complete Deep GLR pipeline, showing the integration of CNN feature learning with analytical PFBS optimization.

\begin{algorithm}[!t]
\caption{Deep GLR CT Reconstruction}
\label{alg:deep_glr}
\begin{algorithmic}[1]
\REQUIRE Sinogram $\mathbf{y}$
\REQUIRE CNN weights $\theta_{Yb}, \theta_F, \theta_\mu$  
\REQUIRE Learnable $\alpha \in [0, 0.1], \varepsilon \in [1.0, 1.5]$
\ENSURE Reconstructed image $\mathbf{x}_{\text{final}}$
\STATE $\mathbf{x}^0 \leftarrow \text{FBP}(\mathbf{y})$
\FOR{$k = 0$ \textbf{to} $39$}
    \STATE $\tilde{\mathbf{x}} \leftarrow \text{CNN}_{Yb}(\mathbf{x}^k)$ \COMMENT{Pre-filtering}
    \STATE $\mathbf{f} \leftarrow \text{CNN}_F(\tilde{\mathbf{x}})$ \COMMENT{Feature extraction}
    \STATE $\mu \leftarrow \text{CNN}_\mu(\tilde{\mathbf{x}})$ \COMMENT{Regularization weight}
    \STATE Build 8-connected graph with weights $w_{ij}~=~\exp(-\|\mathbf{f}_i - \mathbf{f}_j\|_2^2/(2\varepsilon^2))$
    \STATE $\mathbf{L} \leftarrow \mathbf{D} - \mathbf{W}$ \COMMENT{Graph Laplacian}
    \STATE $\nabla_{\text{data}} \leftarrow 2\mathbf{A}^T(\mathbf{A}\mathbf{x}^k - \mathbf{y})$ \COMMENT{Data fidelity gradient}
    \STATE $\nabla_{\text{GLR}} \leftarrow 2\mathbf{L}\mathbf{x}^k$ \COMMENT{GLR gradient}
    \STATE $\mathbf{x}_{\text{temp}} \leftarrow \mathbf{x}^k - \alpha(\nabla_{\text{data}} + \mu\nabla_{\text{GLR}})$ \COMMENT{PFBS update}
    \STATE $\mathbf{x}^{k+1} \leftarrow \mathbf{x}_{\text{temp}} + 0.1\mathbf{x}^k$ \COMMENT{Residual connection}
\ENDFOR
\STATE $\mathbf{x}_{\text{final}} \leftarrow \text{PostProcessing}(\mathbf{x}^{40})$ \COMMENT{Final refinement}
\RETURN $\mathbf{x}_{\text{final}}$
\end{algorithmic}
\end{algorithm}

Algorithm~\ref{alg:deep_glr} details the complete Deep GLR reconstruction process, illustrating the transformation of an image through the complete CNN and PFBS framework.

\section{Experimental Setup}

\subsection{Dataset and Evaluation Protocol}
Experiments employ the LoDoPaB-CT benchmark dataset~\cite{leuschner2021lodopab}, which provides standardized low-dose CT reconstruction evaluation. Following Leuschner et al.'s specifications:
\begin{itemize}
\item Image resolution: 362$\times$362 pixels on 26$\times$26 cm physical domain
\item Projection geometry: Parallel beam with 1000 equidistant angles over $[0,\pi]$ and 513 detector bins
\item Noise model: Poisson statistics with $N_0 = 4096$ mean photons per detector bin
\item Ground truth: High-quality reconstructions from Lung Image Database Consortium/Image Database Resource Initiative (LIDC/IDRI) lung CT database
\item Training subset: 1000 samples representing 2.8\% of full training set (35,820 samples)
\item Validation subset: 100 samples
\end{itemize}

PSNR calculation follows Leuschner et al.'s recommendation for medical imaging applications~\cite{leuschner2021lodopab}:
\begin{equation}
\text{PSNR} = 10 \log_{10}\left(\frac{(\max(\mathbf{x}) - \min(\mathbf{x}))^2}{\text{MSE}}\right)
\label{eq:psnr}
\end{equation}

This dynamic range calculation is preferred over fixed maximum values since the reference value of 3071 Hounsfield Units (HU) used in standard calculations is far from typical CT intensity ranges.

\subsection{Training Protocol}

 Model training employs a two-stage adaptive learning approach with early stopping for computational efficiency. Stage 1 utilizes learning rate $2 \times 10^{-4}$ with cosine annealing and stops at epoch 14 due to validation loss convergence. Stage 2 continues with reduced learning rate $1 \times 10^{-5}$ for fine-tuning and stops at epoch 16, resulting in 30 total training epochs.

The training configuration uses Adam optimizer with weight decay $1 \times 10^{-5}$ and batch size 1 to accommodate full 362$\times$362 image processing without patch-based segmentation that could introduce boundary artifacts. The loss function combines mean squared error with regularization terms for the learnable parameters $\varepsilon$ and $\mu$, ensuring stable parameter evolution within their respective constraint ranges.

\subsection{Implementation Details}
The framework employs the Operator Discretization Library (ODL) for CT operations following LoDoPaB specifications. The implementation uses parallel beam geometry with image domain $[-13.0, 13.0]^2$ cm and Hann filtering with frequency scaling 0.641 for FBP operations. Memory constraints necessitate batch size 1, enabling full-image processing while maintaining computational feasibility.

\section{Results}

\subsection{Quantitative Performance and Data Efficiency Analysis}
Table~\ref{tab:performance} presents reconstruction performance on the LoDoPaB validation subset, demonstrating Deep GLR effectiveness with limited training data.

\begin{table}[!t]
\centering
\caption{Performance Comparison on LoDoPaB-CT Dataset}
\label{tab:performance}
\resizebox{\columnwidth}{!}{%
\begin{tabular}{lccccc}
\toprule
Method & PSNR (dB) & SSIM & Parameters & Training Data & Epochs \\
\midrule
FBP & 24.37 & 0.6916 & -- & -- & -- \\
\textbf{Deep GLR} & \textbf{30.70} & \textbf{0.7719} & \textbf{91,848} & \textbf{1000 samples} & \textbf{30} \\
FBP+U-Net$^{*}$ & 35.84 & 0.8443 & $\sim$500,000 & 35,820 samples & 250 \\
VQ-CAE$^{**}$ & 43.75 & 0.96 & $\sim$500,000 & Full dataset & 250 \\
\bottomrule
\end{tabular}%
}
\begin{tablenotes}
\item[$*$] * Results from LoDoPaB benchmark paper~\cite{leuschner2021lodopab}
\item[$**$] ** Vector Quantized Convolutional Autoencoder from Ramanathan \& Ramasundaram~\cite{ramanathan2023low}
\end{tablenotes}
\end{table}

Deep GLR achieves 6.33 dB improvement over FBP baseline while utilizing only 1000 training samples, representing 2.8\% of the full LoDoPaB training set used by benchmark methods. This data efficiency demonstrates the method's effectiveness with limited training data, an advantage particularly relevant for medical imaging where large annotated datasets may not be readily available due to privacy constraints and annotation costs. The 91,848 parameter count constitutes approximately one-fifth of competing deep learning approaches.

\begin{figure*}[!t]
\centering
\includegraphics[width=0.85\linewidth]{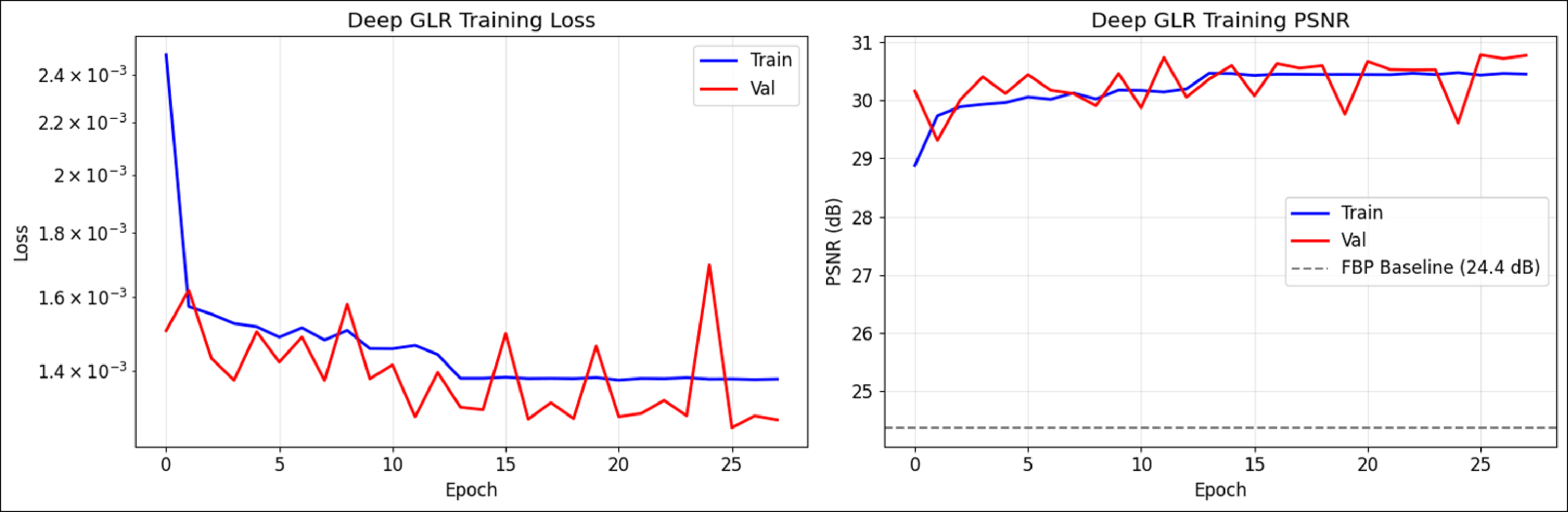}
    \caption{Training loss and PSNR progression over 30 epochs showing stable convergence with early stopping. The curves demonstrate effective learning without overfitting on the limited training dataset}
\label{fig:training}
\end{figure*}

\subsection{Training Dynamics and Parameter Stability}
Fig.~\ref{fig:training} illustrates the training progression over 30 epochs, showing both loss reduction and PSNR improvement. The training curves demonstrate stable convergence without overfitting, with validation PSNR reaching 30.77 dB at completion across both training stages.

Parameter evolution during training reveals stable optimization behavior according to the training logs. The learnable epsilon parameter converged to 1.25 across training epochs, positioning itself at the center of the constrained range $[1.0, 1.5]$. The adaptive regularization parameters $\mu$ learned by CNN$_\mu$ remained within the designed range, demonstrating the model's ability to adjust regularization strength based on input characteristics.

The stable parameter evolution throughout training validates the mathematical framework's robustness and suggests that the learned parameters capture meaningful relationships in CT image reconstruction rather than memorizing training-specific patterns.

\begin{figure*}[t]
\centering
\includegraphics[width=0.8\textwidth]{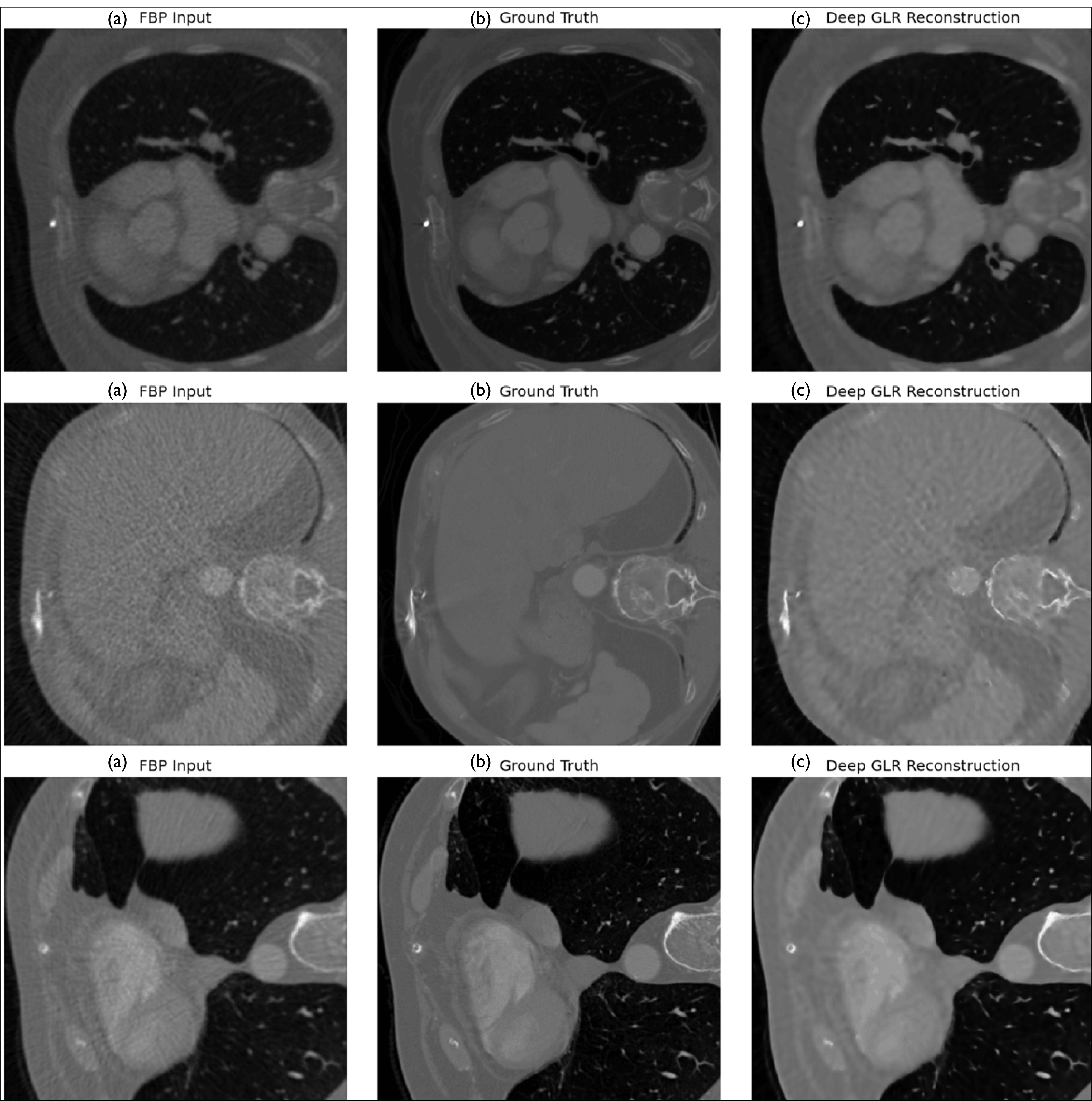}
\caption{Representative reconstruction results on LoDoPaB validation data. Three patient cases showing (a) FBP input with noise artifacts, (b) ground truth reference, and (c) Deep GLR reconstruction. Each case demonstrates consistent noise reduction with anatomical structure preservation across different patient anatomy from the thoracic CT dataset}
\label{fig:results}
\end{figure*}

\subsection{Visual Quality Assessment and Reconstruction Examples}
Fig.~\ref{fig:results} presents representative reconstruction results across three patient cases demonstrating Deep GLR's noise reduction capabilities while maintaining diagnostic image quality. The comparison between FBP input, ground truth, and Deep GLR reconstruction reveals substantial noise artifact elimination without loss of anatomical detail.

These visual results confirm that the graph-based regularization adapts to different anatomical regions represented in the validation dataset without requiring region-specific parameter tuning. 

\section{Discussion}



\subsection{Efficiency-Quality Tradeoffs}

The experimental results demonstrate that parameter-efficient graph 
regularization achieves meaningful noise reduction despite using 5.8 times fewer parameters per dB 
improvement compared to FBP+U-Net, consistent with 
   recent findings that parameter-efficient methods can achieve substantial 
   performance gains in low-data medical imaging scenarios \cite{dutt2024parameter}. This efficiency advantage stems from 
three architectural decisions:

First, the Graph Laplacian's quadratic formulation enables gradient 
computation in $O(n)$ time without iterative thresholding operations 
required by $L_1$-norm methods. This computational efficiency translates to 
both faster training iterations and reduced parameter requirements for 
achieving stable gradients.

Second, parameter sharing across 40 PFBS iterations effectively amortizes 
learning capacity \cite{monga2021algorithm}, the same 91,848 parameters refine the reconstruction 40 
times, multiplying representational power without increasing model size. 
This architectural prior assumes iterative refinement follows consistent 
principles across stages.

Third, integration with classical PFBS optimization provides mathematical 
inductive bias through explicit data fidelity and regularization terms. 
This structure reduces the learning burden compared to pure deep learning 
approaches that must discover both forward model physics and image priors 
entirely from data.

However, this efficiency comes with a performance cost: the 13.05 dB gap 
versus VQ-CAE (43.75 dB) reflects limitations of quadratic 
regularization, which prioritizes smoothness over edge preservation. We observe a sacrifice of performance for 
deployability.

\subsection{Data Efficiency and Medical Imaging Implications}

Training on 1000 samples while achieving better performance than FBP+U-Net has specific implications for 
medical imaging, where annotation-efficient 
   methods have demonstrated that careful regularization enables comparable 
   results with 10\% of typical training data \cite{zhou2021annotation}.  The data efficiency enables three deployment scenarios: (1) federated learning applications where institutions train local models without sharing patient data, (2) rare pathologies where collecting large datasets is infeasible, and (3) emerging healthcare systems lacking extensive imaging databases, allowing faster initial deployment.

The graph-based regularization appears to effectively compensate for 
limited training data by imposing strong structural priors (local 
smoothness). However, this hypothesis requires validation across diverse 
anatomical regions and pathologies beyond the LoDoPaB thoracic CT dataset.

\subsection{Learned Parameter Interpretability}

The convergence of $\epsilon$ to 1.25 provides insight into learned 
representations. Pang et al. \cite{pang2017graph} demonstrated that graph Laplacian regularizers converge to 
   continuous-domain functionals integrating gradient norms in adaptive metric 
   spaces. For CT images with sharp 
tissue boundaries (e.g., bone-soft tissue interfaces), this threshold 
effectively segments homogeneous regions while preserving major anatomical 
boundaries.

The image-adaptive $\mu$ values learned by CNN$_{\mu}$ suggest the network discovered 
that different image regions require different regularization strengths - higher for noisy regions, lower for high-contrast structures. This 
adaptivity emerges from training without explicit supervision, indicating 
the framework can discover problem-appropriate regularization strategies.

\subsection{Limitations and Fundamental Tradeoffs}

The 2D slice-wise processing approach processes each CT slice 
independently, missing potential volumetric consistency information. Real 
clinical protocols acquire 3D volumes where inter-slice relationships 
could improve reconstruction - a limitation of the current implementation. The quadratic regularization formulation may also over-smoothe fine anatomical structures resulting in potential detail loss in complex structures.

In addition, evaluation on a single dataset limits 
generalization abilities. Real clinical scanners employ helical cone-beam 
geometries with different noise characteristics, scatter effects, and 
metal artifacts. Method performance under these conditions remains 
unknown.

The 30-epoch training duration, while demonstrating rapid convergence, 
leaves open whether extended training would close the performance gap 
versus benchmark methods or whether architectural constraints impose 
fundamental quality ceilings.

\subsection{Clinical Readiness Assessment}

While 30.70 dB represents substantial noise reduction over FBP (24.37 dB), 
clinical deployment requires validation beyond PSNR metrics. Radiologist 
evaluation for diagnostic sufficiency, task-specific assessment (lesion 
detection sensitivity/specificity), and testing on real scanner data with 
clinical protocols remain necessary before clinical translation.

This work demonstrates feasibility of parameter-efficient reconstruction 
but does not claim clinical readiness. The method provides a baseline 
showing that graph-based regularization can achieve acceptable quality 
under resource constraints, which is a potential first step toward practical 
deployment in resource-limited settings.

\section{Future Work}

This work demonstrates efficiency-quality tradeoffs over achieving maximum reconstruction performance, with several limitations suggesting future research directions.

Evaluation on a single dataset (LoDoPaB-CT parallel beam geometry) limits generalization claims. Clinical translation requires validation across diverse anatomical regions, imaging protocols, and scanner geometries.

The 1000-sample training constraint and 92K parameter budget demonstrate resource efficiency but impose a 13~dB performance gap versus state-of-the-art methods. Future work should systematically evaluate scaling relationships: at what parameter counts do efficiency gains saturate? What minimum training data yields diagnostically acceptable quality?

Current 2D slice-wise processing misses volumetric consistency present in clinical 3D acquisitions. Extending graph construction to 3D volumes would capture inter-slice relationships but requires addressing computational challenges - 26$\times$ larger Laplacian matrices - potentially through hierarchical graph representations or sparse approximations maintaining efficiency advantages.

\section{Conclusion}
Deep Graph Laplacian Regularization demonstrates effective integration of classical optimization with modern deep learning for low-dose CT reconstruction. The proposed method achieves a 6.33 dB improvement over analytical reconstruction while using only 91,848 parameters and training on just 2.8\% of the standard dataset size. The quadratic GLR formulation provides both computational stability and mathematical interpretability through explicit graph-based modeling.

These results indicate that Deep GLR offers a practical and efficient solution for medical imaging applications where parameter efficiency and data efficiency are critical constraints. Future work will explore scaling the approach to larger datasets and extending the framework to 3D CT reconstruction.

\end{document}